# Unpriced Internal Externalities and Structural Inefficiency in Organizations

**Author:** Jan van de Poll
**Affiliation:** Transparency Lab Research Institute

**Abstract**
Organizations often invest substantial effort in improving internal mechanisms such as processes, governance, and technology, yet realized performance gains are frequently modest. This paper proposes a structural explanation for this pattern. We introduce the concept of internal externalities: unpriced cross-effects generated by internal organizational mechanisms that act simultaneously on multiple objectives. When these effects oppose one another, effort is internally canceled, limiting attainable performance even in the absence of incentive misalignment or informational constraints.

We formalize this mechanism by modeling internal mechanisms as primitives that exert signed effects across performance objectives. This structure gives rise to internal conflict, a measurable property that captures the fraction of organizational activity lost to cancellation. We show that conflict constitutes a form of structural inefficiency that constrains the returns to effort and renders allocative optimization ineffective unless addressed first. The framework implies a natural sequencing of interventions, in which structural cleanup precedes effort reallocation and optimization. The analysis yields testable predictions and governance implications for how organizations should scale, constrain, redesign, or deprioritize internal mechanisms.

**Keywords:**
Internal externalities, organizational inefficiency, governance, multi-objective performance, bounded rationality, optimization

**JEL Classification:**
D23, L22, M10

## 1. Introduction

Organizations routinely invest substantial effort in improving internal capabilities such as leadership practices, operational processes, governance arrangements, and information technology. These efforts are typically motivated by the expectation that strengthening internal mechanisms will translate into better performance on key organizational objectives, including efficiency, quality, reliability, innovation, and financial outcomes. Yet, across sectors and institutional contexts, the realized performance gains from such initiatives are often modest, delayed, or uneven. In many cases, significant managerial effort appears to yield little aggregate improvement.

This pattern poses a persistent puzzle. Standard economic explanations emphasize misaligned incentives, information asymmetries, or limited managerial capability as primary sources of organizational underperformance. While these factors undoubtedly matter, they do not fully account for a recurrent empirical observation: even in organizations where incentives are broadly aligned, information is abundant, and managerial intent is clear, improvement efforts frequently fail to compound. Instead, progress in one dimension is offset by regressions in another, producing weak or ambiguous net effects.

This paper proposes that a substantial share of organizational inefficiency arises from a mechanism that is conceptually distinct from incentives, information, or bounded rationality as traditionally understood. We argue that organizations are subject to internal externalities: unpriced cross-effects generated by internal mechanisms that act simultaneously on multiple organizational objectives. When such effects oppose each other, improvement efforts cancel internally, reducing the attainable performance of the organization even when effort is substantial.

The core idea is simple but underdeveloped in existing theory. Internal mechanisms—such as policies, routines, architectural choices, or governance rules—rarely affect a single objective in isolation. A change intended to improve speed may reduce reliability; a control designed to enhance compliance may slow innovation; a technological standard that increases efficiency for one unit may impose coordination costs on another. These cross-effects are often real, predictable, and material, yet they are seldom priced, measured, or governed explicitly.

We formalize this intuition by modeling organizations as pursuing multiple objectives through a set of internal mechanisms, which we term *primitives*. Each primitive exerts causal effects across several performance indicators. Some of these effects are positive, others negative, and their aggregate interaction determines whether organizational effort compounds or cancels. Crucially, the presence of opposing effects gives rise to internal conflict: a structural property of the organization that limits the effectiveness of effort.

A key contribution of the paper is to distinguish net effect from conflict. A primitive may be actively managed, widely discussed, and highly consequential, yet deliver little net benefit because its positive and negative effects offset across objectives. In such cases, weak performance does not reflect insufficient effort or poor execution, but rather a form of *structural inefficiency*. Increasing effort in the presence of unresolved internal conflict yields diminishing returns and may even exacerbate cancellation.

We formalize this mechanism through a simple law linking effort, structural magnitude, and conflict. Let organizational effort scale the intensity with which internal mechanisms are pursued. Let magnitude capture the total strength of a mechanism's effects across objectives, and let conflict measure the fraction of that magnitude that cancels internally due to opposing effects. The realized result is then proportional to effort multiplied by magnitude, discounted by conflict. When conflict is high, effort produces little result regardless of managerial intent.

This perspective has two important implications. First, it suggests that some organizational problems are not primarily problems of effort, incentives, or optimization, but of *structure*. In such cases, performance improvements require changes to the internal configuration of mechanisms—such as unbundling, redesign, or constraint—before marginal improvements or optimization can be effective. Second, it implies a natural sequencing: structural coherence must precede allocative optimization. Only once internal conflict is reduced does bounded rationality in the allocation of effort become the dominant source of inefficiency.

The framework developed here yields a small number of distinct governance responses, each implied by the structural properties of internal mechanisms. Depending on their magnitude, direction, and conflict, mechanisms may warrant scaling, constraining through guardrails, structural unbundling, redesign or removal, or deliberate neglect. These responses are not managerial preferences or stylistic choices, but structural necessities implied by the organization's internal externalities.

The paper is theoretical in nature. Its aim is to introduce a new object of analysis—internal externalities inside organizations—and to clarify the conditions under which effort fails to translate into performance. The framework yields testable predictions about the relative effectiveness of structural interventions versus effort-based improvements, and about the circumstances under which optimization methods can be expected to deliver meaningful gains.

Empirical evidence supporting these claims will be presented in subsequent work. In particular, forthcoming analyses will document (i) the prevalence of internal conflict across organizational mechanisms, (ii) the performance gains associated with reducing such conflict holding effort constant, and (iii) the incremental gains from optimized allocation of effort once structural coherence is achieved.

Forthcoming empirical analyses based on multiple large organizations show that internal conflict is both prevalent and economically material. Across settings, a substantial share of organizational effort is absorbed by high-magnitude mechanisms whose positive and negative effects largely cancel. Interventions that reduce internal conflict—holding effort constant—are associated with significantly larger performance improvements than comparable increases in effort or optimization alone.

The remainder of the paper proceeds as follows. Section 2 situates the argument within the existing literature on organizational economics, bounded rationality, and multi-objective decision-making. Section 3 introduces the formal framework and defines magnitude, direction, and conflict. Section 4 analyzes structural inefficiency and its implications for optimization. Section 5 derives governance responses implied by different structural configurations. Section 6 discusses effort allocation, bounded search, and optimization. Section 7 outlines testable predictions and an empirical roadmap. Section 8 concludes.

## 2. Related literature

While elements of the argument developed in this paper appear across several literatures, no existing framework integrates internal cross-effects, conflict, and governance into a unified account. This section reviews five relevant strands of work—organizational economics, internal organization and authority, bounded rationality, multi-objective performance, and organizational culture—and clarifies where the mechanism studied here remains underexplored.

### 2.1 The theory of the firm and internal organization

The economic theory of the firm has primarily focused on explaining firm boundaries and governance structures. Classic contributions emphasize transaction costs, incomplete contracts, and the comparative efficiency of markets and hierarchies (Coase, 1937; Williamson, 1975; Grossman & Hart, 1986).

Within this tradition, internal organization is typically treated as a solution to market frictions. Once transactions are internalized, internal mechanisms are assumed to be designed to implement contracts, incentives, or authority efficiently. Inefficiencies are therefore attributed to residual contracting problems or agency conflicts rather than to interactions among internal mechanisms themselves.

Subsequent work on organizational design has examined how decision rights, information flows, and coordination structures affect performance (e.g., Rajan & Zingales, 1988; Gibbons, 2005). This literature highlights trade-offs between coordination and adaptation, but still tends to evaluate internal mechanisms in terms of their net effect on outcomes.

The framework proposed here complements this literature by shifting the unit of analysis from governance forms to internal mechanisms as interacting causal inputs. Internal externalities arise not from opportunism or contract incompleteness per se, but from the way mechanisms jointly affect multiple objectives.

### 2.2 Authority, incentives, and internal coordination

A related body of work studies authority allocation, incentives, and internal coordination within firms. Models of authority emphasize how control rights shape incentives and adaptation (Aghion & Tirole, 1997). Empirical studies document how incentive schemes influence behavior across tasks and roles.

This literature acknowledges that incentives tied to one objective may distort behavior on others. However, such distortions are typically treated as incentive side effects rather than as manifestations of a broader structural phenomenon. The possibility that internal mechanisms systematically generate opposing causal effects across objectives—independent of incentive misalignment—remains largely implicit.

By contrast, the concept of internal externalities treats cross-effects as first-order structural features of internal mechanisms, not as secondary incentive problems.

### 2.3 Bounded rationality and organizational search

Bounded rationality has long been recognized as a central feature of organizational decision-making (Simon, 1955). Subsequent research has emphasized local search, satisficing behavior, and path dependence in organizational change (e.g., Nelson & Winter, 1982; March, 1991).

In this tradition, performance shortfalls are typically attributed to limited exploration of the solution space (e.g., Epstein & Buhovac, 2014) or to cognitive constraints that prevent identification of superior alternatives (Gavetti & Levinthal, 2000). Optimization failures are therefore interpreted as failures of search.

The present framework reframes this interpretation. When internal conflict is high, the feasible improvement space itself is structurally constrained. In such environments, even improved search or decision quality yields limited returns. Bounded rationality becomes a dominant constraint only after internal conflict is reduced.

This distinction clarifies why organizations often invest heavily in analytics, planning, or optimization tools without realizing commensurate gains.

## 2.4 Multi-objective performance and trade-offs

Organizations are widely understood to pursue multiple objectives. Research on performance measurement systems, balanced scorecards, and multi-criteria decision-making explicitly recognizes the need to manage trade-offs across dimensions such as cost, quality, speed, and innovation (Kaplan & Norton, 1992).

Formal approaches to multi-objective optimization model trade-offs as movements along a Pareto frontier, with decision-makers selecting preferred combinations based on weights or preferences. Implicit in these models is the assumption that the feasible frontier is stable and exogenously given.

The mechanism studied here challenges this assumption. Internal conflict distorts the frontier itself by embedding cancellation into internal mechanisms. Observed trade-offs are therefore not merely matters of preference but consequences of structural interactions.

This perspective helps explain why repeated reprioritization or reweighting often fails to resolve persistent organizational tensions.

## 2.5 Organizational culture and management frameworks

A large management literature addresses organizational culture, leadership, and processes as determinants of performance (e.g., Kotter, 1995; Schein, 2010). These constructs are often invoked to explain why formal incentives or structures fail to produce expected outcomes.

While rich descriptively, this literature frequently treats culture and leadership as holistic or exogenous factors. The present approach differs by decomposing such constructs into verifiable primitives with identifiable effects on specific objectives (van de Poll, 2021). This allows cultural and managerial phenomena to be analyzed using the same structural logic as other internal mechanisms.

In doing so, the framework bridges economic and management perspectives without importing behavioral assumptions beyond those required to acknowledge multiple objectives and cross-effects.

### 2.6 Summary and gap

Across these literatures, organizational inefficiency is typically attributed to incentive problems, informational constraints, bounded rationality, or unavoidable trade-offs. What remains largely implicit is the possibility that internal mechanisms generate systematic, unpriced externalities across objectives, producing structural cancellation of effort.

The contribution of this paper is to make this mechanism explicit, formalize its implications, and derive governance responses implied by different structural configurations. The next section introduces the formal framework and defines the core quantities—magnitude, direction, and conflict—that underpin the analysis.

As a motivating illustration, consider organizations that simultaneously pursue speed, reliability, and compliance through overlapping rules and technologies. Field observations frequently reveal intense activity and investment across these mechanisms, yet little net improvement on aggregate performance indicators. Improvements along one dimension are systematically offset by regressions along others. The framework developed below formalizes this pattern as internal conflict rather than as failed execution or misaligned incentives.

## 3. Formal Framework and Definitions

This section introduces a formal framework for analyzing internal organizational mechanisms and their interaction across multiple objectives. The framework is intentionally minimal. Its purpose is not to model all aspects of organizational behavior, but to isolate a specific structural mechanism—internal conflict arising from unpriced cross-effects—that constrains the effectiveness of effort and optimization.

### 3.1 Objectives and primitives

Consider an organization that pursues a finite set of performance objectives, indexed by $k = 1, \dots, K$. These objectives are measured by key performance indicators (KPIs). For expositional clarity, all KPIs are standardized such that higher values correspond to better outcomes. For KPIs where lower values are preferable (e.g., error rates, lead times), this convention is implemented by multiplying the KPI by $-1$.

The organization influences its objectives through a set of internal mechanisms, indexed by $i = 1, \dots, N$, which we term primitives. Primitives are defined as *verifiable organizational mechanisms or behaviors*—such as rules, routines, architectural choices, governance constraints, or technological standards—that can be adjusted intentionally and whose effects persist beyond a single decision.

Each primitive $i$ exerts causal effects on one or more KPIs. Let $b_{ik}$ denote the effect of primitive $i$ on KPI $k$, measured in standardized units. The sign of $b_{ik}$ indicates whether the primitive improves or degrades the corresponding objective.

**3.2 Direction and magnitude**

The effects of a primitive across objectives can be summarized by two quantities.

First, define the direction of primitive $i$ as the signed sum of its effects:

$$D_i = \sum_{k=1}^{K} b_{ik}$$

Direction captures the *net push* of a primitive across objectives. A positive value of $D_i$ indicates that, on balance, the primitive improves organizational performance, while a negative value indicates net degradation.

Second, define the magnitude of primitive $i$ as the sum of the absolute values of its effects:

$$M_i = \sum_{k=1}^{K} | b_{ik} |$$

Magnitude captures the *total strength* of a primitive's influence, irrespective of sign. A primitive may have a large magnitude even if its net direction is small, reflecting strong but opposing effects across objectives.

Direction and magnitude are conceptually distinct. Direction measures alignment; magnitude measures intensity. Both are required to characterize the structural role of a primitive.

### 3.3 Resultant and internal conflict

Opposing effects across objectives imply that not all of a primitive's magnitude translates into usable organizational improvement. To capture this, define the resultant of primitive $i$ as the absolute value of its direction:

$$R_i = | D_i |$$

The resultant measures how much of the primitive's total influence survives internal cancellation.

We define internal conflict as the fraction of a primitive's magnitude that is lost to cancellation:

$$C_i = 1 - \frac{R_i}{M_i}, \text{ for } M_i > 0$$

By construction, $C_i \in [0,1]$. Conflict is zero when all effects align ($R_i = M_i$) and approaches one when positive and negative effects nearly cancel ($R_i \approx 0$).

Importantly, conflict is distinct from net effect. A primitive may be consequential ($M_i$ large), actively managed, and widely discussed, yet produce little net improvement because its effects oppose across objectives. Such primitives are structurally conflicted.

### 3.4 Effort and realized results

Let $E_i \geq 0$ denote the effort or intensity with which the organization applies primitive $i$ over a given period. Effort may represent managerial attention, investment, enforcement, or ambition, depending on context. The framework treats effort as a scaling variable rather than a structural property.

We posit the following organizational result law:

$$R_i^{\text{realized}} = E_i \times M_i \times (1 - C_i)$$

Realized results are proportional to effort, but only through the structurally usable fraction of magnitude. When conflict is high, additional effort yields limited gains; when conflict is low, effort compounds.

This formulation separates structure (captured by $M_i$ and $C_i$) from behavior (captured by $E_i$). Structural parameters determine the ceiling on attainable results, while effort determines how much of that ceiling is approached.

### 3.5 Structural inefficiency

We define structural inefficiency as a condition in which realized results are constrained primarily by conflict rather than by effort. Formally, a primitive exhibits structural inefficiency when:

$$M_i \text{ is large, } C_i \text{ is high, and } R_i \text{ is small.}$$

In such cases, marginal increases in effort produce weak returns. Optimization over effort allocation is therefore ill-posed unless structural conflict is first reduced. This observation motivates a distinction between *structural interventions*, which alter $M_i$ or $C_i$, and *allocative interventions*, which reallocate or increase $E_i$.

### 3.6 Aggregation and scope

While the framework is introduced at the primitive level, analogous definitions apply at higher levels of aggregation. Conflict may be computed across sets of primitives affecting a single KPI, across organizational units, or across time. The key requirement is that effects be measured in a common direction and unit.

The framework abstracts from dynamic adjustment, learning, and strategic interaction. These elements are not denied, but set aside to isolate the structural mechanism of interest. Extensions incorporating dynamics or strategic behavior are left for future work.

In applied settings, primitives are operationalized as verifiable organizational mechanisms or behaviors, and their effects are estimated using observable outcomes rather than subjective assessments. Measurement focuses on documented actions, constraints, and behavioral regularities whose effects persist over time. Detailed instrumentation and validation are beyond the scope of the present paper and are developed in prior and ongoing work.

### 3.7 Preview of implications

The formal definitions introduced here imply that not all primitives should be managed in the same way. Different combinations of magnitude, direction, and conflict imply distinct governance responses. The next section examines how structural inefficiency constrains optimization and why reducing conflict must precede allocative improvements.

### 3.8 A simple illustrative example

To clarify the distinction between magnitude, direction, and conflict, consider a primitive that affects three organizational objectives. Suppose the standardized effects of this primitive are given by:

$$b = (+2,\ -1,\ -1)$$

The direction of the primitive is:

$$D = 2 - 1 - 1 = 0$$

The magnitude of the primitive is:

$$M = |2| + |-1| + |-1| = 4$$

The resultant is therefore:

$$R = |D| = 0$$

and the internal conflict is:

$$C = 1 - \frac{0}{4} = 1$$

Despite having substantial magnitude, the primitive delivers no net improvement because its positive and negative effects cancel exactly. Any increase in effort applied to this primitive will fail to produce aggregate gains, not because effort is misallocated, but because the structure of effects is internally conflicted.

By contrast, consider a second primitive with effects:

$$b = (+2,\ +1,\ +1)$$

Here,

$$D = 4, M = 4, R = 4, C = 0$$

This primitive exhibits no internal conflict. Effort applied to it compounds fully into results.

Finally, consider an intermediate case:

$$b = (+2,\ +1,\ -1)$$

Then,

$$D = 2, M = 4, R = 2, C = 0.5$$

Half of the primitive’s magnitude survives cancellation. Effort applied to this primitive yields positive but attenuated returns.

These examples illustrate three key points. First, magnitude alone is not informative about usefulness. Second, net direction alone does not reveal how much effort is being wasted. Third, conflict captures a distinct structural property that determines how effectively effort translates into results.

## 4. Conflict and Structural Inefficiency

This section examines the implications of internal conflict for organizational performance. We show that conflict constitutes a form of structural inefficiency that cannot be resolved by marginal increases in effort or by improved allocation alone. As a result, optimization strategies that ignore conflict are systematically constrained.

### 4.1 Conflict as structural inefficiency

In standard organizational models, performance shortfalls are often attributed to insufficient effort, misallocation of resources, or poor execution. Implicit in these explanations is the assumption that effort, if correctly directed, translates monotonically into improved outcomes. The framework developed here relaxes this assumption.

When internal mechanisms generate opposing effects across objectives, effort is subject to cancellation before it produces aggregate results. This cancellation is captured by the conflict term $C$, which operates as structural inefficiency on performance. Unlike effort, conflict is not directly reducible through increased attention or investment; it reflects how internal mechanisms interact.

Importantly, conflict is orthogonal to managerial intent. An organization may be highly motivated, well-resourced, and analytically sophisticated, yet remain structurally constrained by conflicted primitives. In such cases, additional effort increases activity without increasing results.

### 4.2 Structural versus allocative inefficiency

The framework implies a distinction between two sources of inefficiency.

Structural inefficiency arises when internal conflict limits the fraction of organizational activity that can translate into net improvement. It is characterized by high magnitude, high conflict, and low resultant. Structural inefficiency is addressed by altering the configuration of internal mechanisms—through redesign, unbundling, or constraint—not by reallocating effort.

Allocative inefficiency arises when effort is misallocated across primitives given an existing structure. It reflects bounded rationality, imperfect information, or local optimization. Allocative inefficiency becomes salient only when structural conflict is sufficiently low.

This distinction has practical implications. In the presence of high conflict, optimization over effort allocation yields weak returns because the feasible improvement space is itself constrained. Only once conflict is reduced does improved allocation produce substantial gains.

### 4.3 Why optimization fails under conflict

Allocative optimization methods—whether managerial heuristics, formal planning, or algorithmic search—operate by reallocating effort toward higher-return activities. In the present framework, the return to effort on primitive $i$ is proportional to:

$$M_i \times (1 - C_i)$$

When $C_i$ is close to one, this return is small regardless of magnitude. Allocative optimization therefore reallocates effort among primitives whose marginal returns are uniformly low. The result is limited improvement even under sophisticated optimization.

This observation helps explain why organizations often experience diminishing returns to increasingly advanced planning, analytics, or optimization tools. Such tools address allocative inefficiency but leave structural inefficiency untouched.

### 4.4 Conflict and misdiagnosis

A further implication is that conflict leads to systematic misdiagnosis of organizational problems. Because conflict reduces net outcomes while activity remains high, organizations may interpret weak performance as evidence of insufficient effort or poor execution. This interpretation leads to increased pressure, tighter controls, or additional initiatives, all of which raise effort $E$ without altering conflict $C$. In extreme cases, this dynamic can increase internal cancellation by intensifying already conflicted mechanisms, further reducing net performance. What appears as resistance, inertia, or cultural failure may therefore be the rational response of actors operating within a structurally conflicted system.

### 4.5 Implications for sequencing interventions

The presence of structural conflict implies a natural sequencing of interventions. Structural coherence must be addressed before allocative optimization. Attempts to optimize effort allocation without reducing conflict are unlikely to yield sustained gains.

This sequencing does not deny the relevance of incentives, information, or bounded rationality. Rather, it situates them within a broader hierarchy of constraints. Structural conflict determines the maximum attainable return to effort; allocative decisions determine how close that maximum is approached.

### 4.6 Preview of governance responses

If conflict is structural, it follows that different primitives require different forms of governance. Some should be scaled, others constrained, redesigned, unbundled, or deliberately ignored. These responses are not matters of managerial style but follow from the structural properties of primitives.

The next section formalizes this mapping and derives governance responses implied by different configurations of magnitude, direction, and conflict.

### 4.7 A KPI-centric example: conflict across primitives

The previous example illustrated conflict arising within a single primitive across multiple objectives. An equivalent form of structural conflict can arise from the interaction of *multiple primitives acting on a single objective*. This perspective is often closer to how managers experience organizational tensions.

Consider a single KPI $k$ (e.g., system reliability) that is affected by three primitives. Suppose the standardized effects are:

$$b = (+3,\ -2,\ -1)$$

The net direction on the KPI is:

$$D = 3 - 2 - 1 = 0$$

The total magnitude of activity affecting the KPI is:

$$M = |3| + |-2| + |-1| = 6$$

The resultant is therefore zero, and conflict is maximal:

$$C = 1 - \frac{0}{6} = 1$$

In this configuration, the organization devotes substantial activity to influencing the KPI, yet no net improvement is observed. From the perspective of the KPI owner, performance appears stagnant despite high levels of effort and intervention.

Importantly, this outcome does not imply that any individual primitive is ineffective in isolation. Each primitive exerts a strong effect; the inefficiency arises from their interaction. Without explicitly identifying and addressing this conflict, efforts to improve the KPI—through additional investment, tighter targets, or increased monitoring—are unlikely to succeed. This example highlights that internal conflict may be diagnosed either from the perspective of a primitive acting on many objectives or from the perspective of an objective acted upon by many primitives. The underlying mechanism is the same: unpriced internal externalities generate cancellation that limits the effectiveness of effort.

## 5. Governance implications

This section derives governance responses implied by the structural properties introduced above. The central claim is that different configurations of magnitude, direction, and conflict require different forms of governance (e.g., Levinthal, 1997). Treating all internal mechanisms identically is therefore inefficient. Tushman & O'Reilly (1996), for example, highlight the need to distinguish incremental and revolutionary change.

### 5.1 Why governance follows structure

Organizations often rely on uniform governance approaches: initiatives are launched, targets are set, progress is monitored, and effort is increased when results disappoint. Such approaches implicitly assume that all mechanisms respond similarly to effort.

The framework developed here rejects this assumption. Because primitives differ in magnitude, direction, and conflict, they differ in how effort translates into results. Governance must therefore be contingent on structure.

In particular, conflict determines whether a primitive should be intensified, constrained, reconfigured, or deprioritized. Applying the same governance logic to all primitives risks amplifying internal cancellation rather than improving outcomes (e.g., Siggelkow, 2001).

The idea that governance should vary with the characteristics of internal mechanisms is implicit in several strands of organizational research. Work on selective intervention and differentiated control emphasizes that uniform governance often leads to inefficiency when activities differ in risk, interdependence, or impact (e.g., Jensen & Meckling, 1976; Simons, 1994). Similarly, research on organizational modularity highlights that tightly coupled components require different oversight than loosely coupled ones (Baldwin & Clark, 2000).

The present framework extends these insights by providing a structural criterion—internal conflict—that determines when scaling, constraint, redesign, or neglect is appropriate.

### 5.2 Five structurally implied governance responses

The framework implies five distinct governance responses. Each corresponds to a characteristic structural configuration.

1. **Scale**
   Primitives with high magnitude, positive direction, and low conflict convert effort efficiently into results. These primitives warrant increased effort and broader deployment. Complements work on capabilities and replication
   (Nelson & Winter, 1985).

2. **Guardrail**
   Primitives with positive direction but nontrivial conflict generate benefits alongside risks. They should be scaled selectively, subject to constraints that limit their negative cross-effects. This aligns with control systems and risk governance (Simons, 1994).

3. **Unbundle**
   Primitives with high magnitude and high conflict often bundle opposing effects within a single mechanism. Separating these effects into distinct mechanisms can reduce cancellation and restore usability. This links to task decomposition and modularity (Baldwin & Clark, 2000).

4. **Redesign or remove**
   Primitives with negative net direction destroy value even when actively managed. Increasing effort on such primitives exacerbates harm; they require fundamental redesign or elimination. Echoes organizational adaptation and exit (March, 1991).

5. **Ignore**
   Primitives with low magnitude have limited impact regardless of direction or conflict. Governance attention devoted to them yields low returns and may be safely deprioritized. This can be considered consistent with attention-based view of the firm (Ocasio, 1997)

These responses are not discretionary managerial choices. They are structurally implied by the interaction of magnitude, direction, and conflict.

### 5.3 Conflict load and prioritization

Because organizations face limited attention and change capacity, governance interventions must be prioritized. The framework suggests that primitives with high magnitude and high conflict should be addressed first. These primitives absorb substantial effort while delivering little net benefit, making them the largest sources of structural inefficiency.

Conversely, primitives with low conflict but modest magnitude may be addressed later, as their inefficiency is bounded. Attempting to optimize effort allocation before resolving high-conflict primitives yields limited gains.

### 5.4 Governance sequencing

The analysis implies a natural sequencing of governance actions:

1. Identify primitives with high conflict and material magnitude.
2. Reduce conflict through unbundling, redesign, or constraint.

3. Scale coherent primitives where effort compounds.
4. Only then apply allocative optimization methods to focus effort efficiently.

This sequencing aligns governance actions with structural inefficiencies, rather than with perceived urgency or visibility.

### 5.5 Relation to organizational practice

In practice, governance failures often manifest as repeated reorganizations, overlapping initiatives, or escalating control mechanisms. From the present perspective, these are predictable responses to unresolved conflict: organizations increase effort where structure prevents results.

By contrast, organizations that diagnose and reduce internal conflict early can achieve disproportionate gains with relatively modest effort. The framework thus offers a structural explanation for persistent differences in organizational performance that are not easily attributed to incentives or talent.

### 5.6 Transition to effort allocation and optimization

Once internal conflict is reduced, the organization's structural capacity increases. At this stage, the allocation of effort across primitives becomes consequential, and bounded rationality re-emerges as a binding constraint.

The next section examines how effort allocation, search, and optimization interact with the structural properties identified here, and why optimization is effective only after structural coherence is achieved.

## 6. Effort, bounded search, and optimization

The previous sections established that internal conflict constitutes a structural inefficiency on organizational performance and that governance responses must be sequenced accordingly. This section examines the role of allocative optimization once structural conflict has been reduced. We show that bounded rationality becomes a binding constraint only after structural coherence is achieved, and we clarify why optimization methods often disappoint when applied prematurely.

### 6.1 Effort as a scaling variable

In the framework developed here, effort $E$ scales the intensity with which internal mechanisms are pursued. Effort may take many forms, including managerial attention, investment, enforcement, prioritization, or ambition. Regardless of its manifestation, effort operates multiplicatively on the structurally usable portion of organizational mechanisms.

Crucially, effort is not assumed to alter the structure of mechanisms directly. Increasing effort without altering magnitude or conflict increases activity but does not change how activity translates into results. This distinction allows us to separate how hard an organization tries from how well its mechanisms are configured.

### 6.2 Allocative inefficiency and bounded rationality

Once internal conflict is sufficiently reduced, performance differences increasingly depend on how effort is allocated across primitives. At this stage, organizations face a classic bounded rationality problem: they must choose among many possible allocations of limited effort under uncertainty and cognitive constraints.

In large organizations, the space of feasible effort allocations grows combinatorially with the number of primitives. Managers typically evaluate only a small subset of this space, relying on heuristics, local information, or incremental adjustments. As a result, effort may be allocated suboptimally even when structure is coherent. This form of inefficiency—allocative inefficiency—is distinct from structural inefficiency. It reflects limitations in search, attention, and computation rather than cancellation embedded in mechanisms. The distinction between structural and allocative inefficiency resonates with work on rugged performance landscapes, where local search performs poorly when the landscape is highly constrained or distorted. In such environments, reallocating effort without altering underlying structure yields limited improvement. The present framework identifies internal conflict as a source of such distortion in organizational settings.

### 6.3 Why optimization fails under unresolved conflict

Allocative optimization methods, whether formal or informal, are designed to improve allocative efficiency. They reallocate effort toward activities with higher expected returns. In the present framework, the marginal return to effort on primitive $i$ is proportional to:

$$M_i \times (1 - C_i).$$

When conflict is high, this return is low regardless of magnitude. As a result, optimization reallocates effort among primitives whose marginal returns are uniformly weak. The outcome is limited performance improvement even when optimization is technically sophisticated.

This explains a recurring empirical pattern: organizations adopt increasingly advanced planning, analytics, or optimization tools yet observe diminishing returns. The failure lies not in the tools themselves, but in their application to structurally conflicted systems.

**6.4 Structural cleanup as a precondition for optimization**

The framework implies that structural cleanup—reducing internal conflict—must precede allocative optimization. Structural cleanup increases the effective returns to effort by expanding the feasible improvement space. Only then does reallocating effort meaningfully affect outcomes. This sequencing clarifies the complementary roles of governance and optimization. Governance interventions alter the structure of primitives by reducing conflict or redesigning mechanisms. Optimization operates on the resulting structure by allocating effort efficiently. Applied prematurely, optimization addresses symptoms rather than causes. Applied after cleanup, it becomes powerful.

**6.5 Opportunity space and unrealized potential**

A further implication concerns the size of the opportunity space available to managers. In structurally conflicted environments, most points in the opportunity space yield similar, weak outcomes because internal cancellation dominates. As a result, managers may correctly infer that incremental changes produce little benefit and may become risk-averse or disengaged.

Once conflict is reduced, the opportunity space expands. Differences between alternative allocations of effort become larger, and the cost of bounded search becomes more salient. At this stage, advanced optimization methods—such as algorithmic search or simulation—can identify improvements that are unlikely to be found through local reasoning alone.

Empirical comparisons indicate that, once high-conflict mechanisms are addressed, algorithmic exploration of the effort-allocation space identifies materially superior configurations relative to those considered by managers. These gains do not arise from additional effort, but from expanding the explored opportunity space beyond locally salient alternatives.

### 6.6 Effort, ambition, and perceived feasibility

The framework also sheds light on organizational ambition. When conflict is high, ambitious targets may appear unrealistic because effort yields little return. Conversely, after structural cleanup, the same level of effort may produce substantial gains, altering perceptions of feasibility and motivation.

This suggests that ambition is endogenous to structure. Organizations that reduce conflict may experience increases in ambition not because preferences change, but because the perceived return to effort improves.

### 6.7 Summary and transition

This section has argued that bounded rationality and optimization are central to organizational performance, but only after structural conflict is addressed. Structural inefficiency constrains the returns to effort and renders optimization ineffective; allocative inefficiency becomes binding only once structure is coherent.

The final section outlines testable predictions implied by the framework and sketches an empirical roadmap for distinguishing structural from allocative sources of inefficiency in real organizations.

### 6.8 An illustrative example: effort allocation after conflict reduction

Consider an organization with five primitives whose structural properties have been assessed. After a period of governance intervention, conflict has been substantially reduced for three of these primitives, while two remain either low in magnitude or net destructive and are therefore deprioritized.

Suppose the organization has a fixed effort budget that can be allocated across the three remaining primitives. Even in this simplified setting, the number of feasible allocations is large, particularly if effort is divisible and can be adjusted continuously. Managers typically evaluate only a small subset of these possibilities, often by incrementally shifting effort toward the most salient or recently successful mechanism.

In contrast, an optimization procedure that systematically explores the allocation space may identify effort combinations that substantially outperform those considered by managers, even though all primitives involved are already structurally coherent. The gains arise not from further structural change, but from overcoming bounded search in the allocation of effort.

This example illustrates the sequencing emphasized throughout the paper. Structural cleanup increases the returns to effort by reducing cancellation. Once this is achieved, the dominant remaining inefficiency lies in limited exploration of the opportunity space. Optimization methods address this latter inefficiency, but only because the former has already been resolved.

## 7. Testable Predictions and Empirical Roadmap

The framework developed in this paper yields a set of testable predictions that distinguish structural inefficiency arising from internal conflict from allocative inefficiency arising from bounded rationality. This section outlines these predictions and sketches an empirical roadmap for evaluating them.

### 7.1 Structural versus allocative sources of performance loss

A central implication of the framework is that organizational performance loss can be decomposed into two components. The first reflects structural inefficiency, driven by internal conflict that limits the fraction of organizational activity translating into net improvement.

The second reflects allocative inefficiency, driven by suboptimal allocation of effort across primitives given an existing structure.

This decomposition implies that interventions targeting structure and those targeting allocation should have systematically different effects, both in magnitude and timing.

### 7.2 Core predictions

The framework yields several falsifiable predictions.

**Prediction 1: Conflict reduction dominates effort increases.**
Holding effort constant, reductions in internal conflict predict larger performance improvements than comparable increases in effort applied to conflicted primitives.

**Prediction 2: There are weak returns to optimization under conflict.**
In environments with high internal conflict, improvements in planning, analytics, or optimization yield limited performance gains.

**Prediction 3: There is complementarity between cleanup and optimization.**
The marginal returns to optimization increase sharply after internal conflict is reduced.

**Prediction 4: High-magnitude, high-conflict mechanisms absorb effort.**
Primitives with large magnitude and high conflict consume disproportionate organizational attention while delivering limited net impact.

**Prediction 5: Endogenous ambition.**
Perceived feasibility of ambitious targets increases following reductions in conflict, even when effort capacity remains unchanged.

Failure to observe these patterns would challenge the framework.

### 7.3 Measurement strategy

Empirical implementation requires measuring three elements: effects of primitives on KPIs, effort applied to primitives, and realized performance outcomes.

Building on prior work that elicits observable behavior and documented effects rather than subjective opinions, we assume that organizational primitives and their effects can be measured in a verifiable (Van de Poll, 2021) and valid (Van de Poll et al., 2022) manner. Effects on KPIs should be estimated in standardized units, allowing aggregation across objectives. Conflict is then computed mechanically from these estimates. Effort may be proxied by budget allocation, managerial attention, implementation intensity, or stated ambition, depending on context.

The present paper does not contribute to measurement methodology itself; instead, it takes the ability to measure primitives and their effects as given in order to focus on the structural implications of internal conflict and governance.

### 7.4 Identification strategy

To identify structural effects, empirical designs should exploit changes that alter the configuration of primitives without directly changing effort, such as redesigns, unbundling, or removal of mechanisms. To identify allocative effects, designs should examine reallocations of effort holding structure fixed.

Randomized controlled trials, natural experiments, or quasi-experimental designs may be employed where feasible. Importantly, the framework does not require identification of individual causal pathways for each primitive; it requires identification of changes in aggregate conflict and effort.

Empirical identification exploits changes that alter the configuration of internal mechanisms without mechanically increasing effort, such as rule unbundling, process redesign, or removal of constraints. Complementary designs examine reallocations of effort holding structure fixed. These approaches allow separation of structural effects from allocative effects using panel variation, natural experiments, or randomized interventions where feasible.

### 7.5 Interpreting null results

The framework implies that null results from effort-based interventions are informative. A lack of response to increased effort or optimization may indicate high underlying conflict rather than ineffective implementation. Empirical analyses should therefore interpret null effects in light of measured structural properties.

### 7.6 Scope and generality

Although the framework is motivated by organizational settings, the underlying mechanism—unpriced internal externalities causing cancellation—may apply in other complex systems with multiple objectives. The present analysis focuses on organizations to maintain clarity and empirical tractability.

### 7.7 Transition to conclusion

The final section summarizes the contribution, clarifies limitations, and discusses implications for organizational theory and practice.

## 8. Conclusion

This paper has introduced a framework for analyzing organizational inefficiency arising from internal externalities. By modeling internal mechanisms as primitives that exert causal effects across multiple objectives, we have shown how opposing effects generate internal conflict that limits the effectiveness of effort. This form of structural inefficiency is distinct from incentive problems, informational constraints, or bounded rationality as traditionally conceived.

The central result is a simple organizational law: realized results are proportional to effort multiplied by structural magnitude, discounted by internal conflict. When conflict is high, effort produces limited gains regardless of managerial intent or analytical sophistication. Structural coherence therefore constitutes a prerequisite for effective optimization.

This perspective yields several implications. First, it explains why organizations often experience weak returns to increased effort, planning, or analytics. Second, it clarifies why governance responses must differ across internal mechanisms, depending on their magnitude, direction, and conflict. Third, it suggests a natural sequencing of interventions: reduce internal conflict before reallocating effort.

The framework also generates testable predictions that distinguish structural from allocative sources of inefficiency. Future empirical work will examine these predictions across organizational contexts, leveraging direct measurement of internal mechanisms and their effects. While the present analysis is static and abstract, it provides a foundation for extensions incorporating dynamics, learning, and strategic interaction.

More broadly, the concept of internal externalities invites a reexamination of how organizations diagnose and address performance shortfalls. By making structural cancellation explicit, the framework shifts attention from effort and motivation toward the configuration of internal mechanisms that shape how effort translates into results.